# Electrostatic charge accumulation versus electrochemical doping in $SrTiO_3$ electric double layer transistors

K. Ueno[1] *, H. Shimotani[2], Y. Iwasa[2,3], M. Kawasaki[1, 3, 4]

1) WPI Advanced Institute for Materials Research (AIMR), Tohoku University, Sendai 980-8577, Japan
2) Quantum-Phase Electronics Center, University of Tokyo, Tokyo 113-8656, Japan
3) CREST, Japan Science and Technology Agency, Tokyo 102-0075, Japan
4) Institute for Materials Research, Tohoku University, Sendai 980-8577, Japan

Abstract

In electric double layer transistors with $SrTiO_3$ single crystals, we found distinct differences between electrostatic charge accumulation and electrochemical reaction depending on bias voltages. In contrast to the reversible electrostatic process below 3.7 V with a maximum sheet charge carrier density, $n_S$, of $10^{14}$ cm$^{-2}$, the electrochemical process causes persistent conduction even after removal of the gate bias above 3.75 V. $n_S$ reached $10^{15}$ cm$^{-2}$ at 5 V, and the electron mobility at 2 K was as large as $10^4$ cm$^2$/Vs. This persistent conduction originates from defect formation within a few micrometers depth of $SrTiO_3$.

* uenok@imr.tohoku.ac.jp

The electric field effect is attracting renewed interest owing to its ability to induce a phase transition from an insulator to a superconductor, by accumulating high-density ($10^{13}$–$10^{14}$ cm$^{-2}$) charge carriers through a capacitively coupled gate bias.[1,2] This has become possible with an electric double layer transistor (EDLT) geometry, which uses an electrolyte as a gate dielectric of a field-effect transistor.[3–8] An electric double layer with a thickness of around 1 nm is formed between the electrolyte and a semiconductor channel, and one can attain electric fields as high as 50 MV/cm.[6,9] The electrolyte can be either a polymer containing supporting salt[1,3–7] or an ionic liquid.[2,8] This technique allows studies of the basic physics and its possible applications.

It is found that a large electric field in the electric double layer induces not only electrostatic charge accumulation but also an electrochemical reaction between the electrolyte and the semiconductor. Electrochemical insertion of ions results in bulk conduction in a semiconductor through chemical doping. Indeed, an electrochemical transistor with the same device configuration has been reported with organic conductive polymers and nanotubes by using their reversible redox reaction.[10,11] This electrochemical doping is a possible origin of the superconductivity in a semiconductor channel in an EDLT. Furthermore, an electrochemical reaction can permanently damage a channel of an EDLT. Takeya *et al.* reported the irreproducibility of organic EDLT device characteristics under excess gate bias, and they have attributed this to the channel's permanent damage caused by the electrochemical reaction.[5] An irreversible change in transport properties by an electrochemical reaction has also been reported for cuprate superconductors and anatase $TiO_2$ thin films.[12,13]

In this paper, we report the device characteristics at high gate biases for an EDLT with a $SrTiO_3$ single crystal channel and a polymer electrolyte that is almost identical to the one used in our previous study.[1] We found that a gate bias range with electrostatic charging was clearly distinguished from that where an electrochemical reaction gave bulk conduction. The superconductivity reported previously is within the regime of electrostatic charge accumulation.[1]

EDLT devices were fabricated on a (001) surface of a $SrTiO_3$ single crystal.[1] Figure 1(a) shows a schematic diagram of the device. All the source/drain and gate electrodes were fabricated by electron-beam evaporation of Au/Ti and a photolithographic patterning. For the Ohmic source/drain contacts, the $SrTiO_3$ surface was reduced by Ar ion milling to form $SrTiO_{3-\delta}$. A hard-baked photoresist layer was used as a separator to isolate the gate electrode from the SrTiO$_3$ channel area. We used a polymer electrolyte, polyethylene oxide (PEO: weight-average molecular weight of 1000) containing KClO$_4$ ([K]:[O] in PEO = 1:100 ). The contact electrodes of the device were fabricated with Hall bar geometry having a 45-μm wide and 250-μm long channel, which allows us to measure the four-terminal resistance and the Hall coefficient. Electrical measurements of the devices were performed with an Agilent 4155C semiconductor analyzer by controlling the ambient temperature and magnetic field by Quantum Design's Physical Properties Measurement System. To facilitate the chemical reaction, switching and transfer characteristics were examined at 320 K, where the electrolyte was in a liquid state.

Figure 1 (b) shows switching characteristics measured at a drain bias, $V_D$, of 0.05 V. Electric voltage pulses with a width of 6 min were repeatedly applied to the gate electrode with a pulse height increasing from 0 to 5 V, as shown in the top panel. Sheet conductance, $\sigma_S$, of the channel (middle panel) was measured using a four-terminal method. For a gate bias, $V_G$, below 3.7 V, $\sigma_S$ rapidly changed its value between a finite value and zero within a second. In contrast, $\sigma_S$ did not return to zero by removing $V_G$ above a threshold of 3.75 V. In addition, a leak current, $I_G$, (bottom panel) exhibited an overshoot with a slow decay upon turning on $V_G$ and a negative value upon removal of $V_G$, in contrast to the fast response of $I_G$ due to the displacement current below the threshold. This complementary $I_G$ behavior with a slow decay is a sign of electrochemical redox reactions at the electrolyte/channel interface, where $I_G$ corresponds to a Faradaic current of an electrochemical reaction. The channel should be reduced by the positive gate bias and oxidized again at the zero gate bias during the decay time. As a result, the semiconductor channel gained a persistent conduction path. It is noted that a single gate sweep to 5 V also caused persistent conduction, indicating that the applied voltage value, rather than the repeated application of gate pulses, is responsible for the electrochemical reaction.[14] By assuming this electrochemical mechanism, we defined an electrochemically induced conduction ($\sigma_{echem}$) as the residual conduction at the off state after applying a certain $V_G$, as shown in the middle panel of Fig. 1(b). An EDL-gated electrostatic conduction ($\sigma_{edl}$) was also defined as a difference in $\sigma_S$ from the off state to the on state.

Figure 2 shows $V_G$ dependence of $\sigma_{edl}$ and $\sigma_{echem}$. $\sigma_{edl}$ exhibited a sharp increase with increasing $V_G$ at around 1.5 V, which corresponds to the threshold voltage of the EDLT. A sharp increase in $\sigma_{echem}$ took place above 3.75 V, which gives a distinct window of electrostatic charge accumulation without electrochemical reaction, as indicated by blue color in Fig. 2. $\sigma_{echem}$ was saturated above 4.3 V, indicating that the electrochemically reacted depth in SrTiO$_3$ did not extend above 4.3 V. In addition, $\sigma_{echem}$ exceeded $\sigma_{edl}$ for $V_G$ above 4 V and was two times larger than $\sigma_{edl}$ for $V_G$ of 5 V. Therefore, conduction at the high gate is dominated by electrochemically induced charge carriers.

It is interesting to compare the transport properties of the channel between the electrostatic and electrochemical regions. After we biased at each $V_G$, the EDLT was cooled during the measurement of a sheet resistance, $R$, and a Hall coefficient, $R_H$. The left panels of Fig. 3 show $R$, a sheet charge carrier density, $n_S = 1/eR_H$, and an electron mobility, $\mu_H$. The right panels of Fig. 3 show $V_G$ dependence of $n_S$ and $\mu_H$ at various temperatures. The channel showed metallic conduction down to 2 K for $V_G$ above 2.0 V. $R_H$ was always negative and $n_S$ was almost temperature independent above 100 K. In the electrostatic region, $n_S$ increased with increasing $V_G$ and saturated to $10^{14}$ cm$^{-2}$ for $V_G$ at around 3.5 V, which is almost the identical dependence given in our previous report.[1] When $V_G$ exceeded 4.0 V, $n_S$ increased again to reach $10^{15}$ cm$^{-2}$ for $V_G$ of 5.0 V. The temperature dependence of $\mu_H$ showed different behaviors in these two regions: $\mu_H$ above 4.0 V is always larger than that below 3.5 V and reaches $10^4$ cm$^2$/Vs at 2 K. This value corresponds to those of the chemically doped bulk single crystals with a volume charge carrier density below $10^{18}$ cm$^{-3}$.[15] By using this volume charge carrier density, the thickness of the conductive region is deduced to be around 10 μm, which is far thicker than that of an EDL-gated accumulation layer of 10 nm.[1]

We also examined microscopic structures of the channel after the electrochemical reaction with a $V_G$ of 5 V by atomic force microscopy (AFM) and transmission electron microscopy (TEM). A sample that experienced only electrostatic charge accumulation showed no appreciable change compared with those of a pristine SrTiO$_3$ single crystal.[1] In contrast, after the electrochemical reaction, the sample showed significant damage, as shown in Fig. 4. An AFM image showed an undulating surface with a peak height of about 10 nm. Note that a step-and-terrace structure seen on the original surface remains on the undulated structure. Therefore, it is considered that some ions or vacancies were intercalated into the SrTiO$_3$ surface and uplifted the surface while maintaining the step-and-terrace structure. A TEM image showed that high-density defects are locally concentrated near the surface, as indicated by a black arrow in Fig. 4 (b), probably corresponding to the undulated structure in the AFM image. In addition, we can find dislocation loops at a depth of several micrometers below the surface, as shown by white arrows in Fig. 4 (b). Therefore, the electrochemical reaction also formed some types of defects in the SrTiO$_3$ matrix with a thickness of several micrometers, probably corresponding to the conductive region found from the transport properties.

Finally, we discuss the origin of the electrochemically induced conduction in the bulk SrTiO$_3$. There are two possible electrochemical reactions: one is an intercalation of K$^+$ cations to SrTiO$_3$ and another is an oxygen removal. We measured the depth profile of K-related species by a secondary ion mass spectrometer for an electrochemically damaged surface. No signal could be seen above the system background. It has been reported that a severe decrease of electron mobility occurs in TiO$_2$ thin films by electrochemical intercalation of Li cations.[16] Therefore, intercalation of cations is not likely for the electrochemically induced conduction. On the other hand, oxygen vacancies in SrTiO$_3$ are known to make a thick conductive layer with an electron mobility of 10$^4$ cm$^2$/Vs at low temperature.[17,18] In addition, the mobility of the oxygen ions is reported to be as large as $4 \times 10^{-12}$ cm$^2$/Vs in (Ba,Sr)TiO$_3$, which is sufficient for producing a 10-μm reduced layer with an acceleration field of 1 MV/cm in several minutes.[19] Therefore, it is plausible to conclude that the conductive layer was formed by the oxygen removal at the surface region of SrTiO$_3$, where the high electric field was applied.

In conclusion, we have demonstrated the clear difference between electrostatic charge accumulation and electrochemical doping in SrTiO$_3$ EDLT devices by using systematic transport measurements at an elevated temperature of 320 K. In contrast to the reversible electrostatic charging below a critical gate voltage of 3.7 V, application of gate voltage above 3.75 V induces persistent conduction, which is attributed to the electrochemical reactions. The large leak current, the low temperature transport properties, and AFM and TEM inspections revealed that the removal of oxygen ions extending to a depth of about 10 μm is responsible for this electrochemical doping. The present result indicates that clarifying the doping mechanism is indispensible to study the uses of EDLT devices.

The authors thank H. T. Yuan for valuable discussions. This study was partly supported by a Grant-in-Aid for Scientific Research (No. 21686002, 21224009, 21654046) from the Ministry of Education, Culture, Sports, Science and Technology of Japan. This study was also partly supported by the Mikiya Science and Technology Foundation, the Murata Science Foundation and the Nippon Sheet Glass Foundation for Materials

Science and Engineering.

# Figure 1

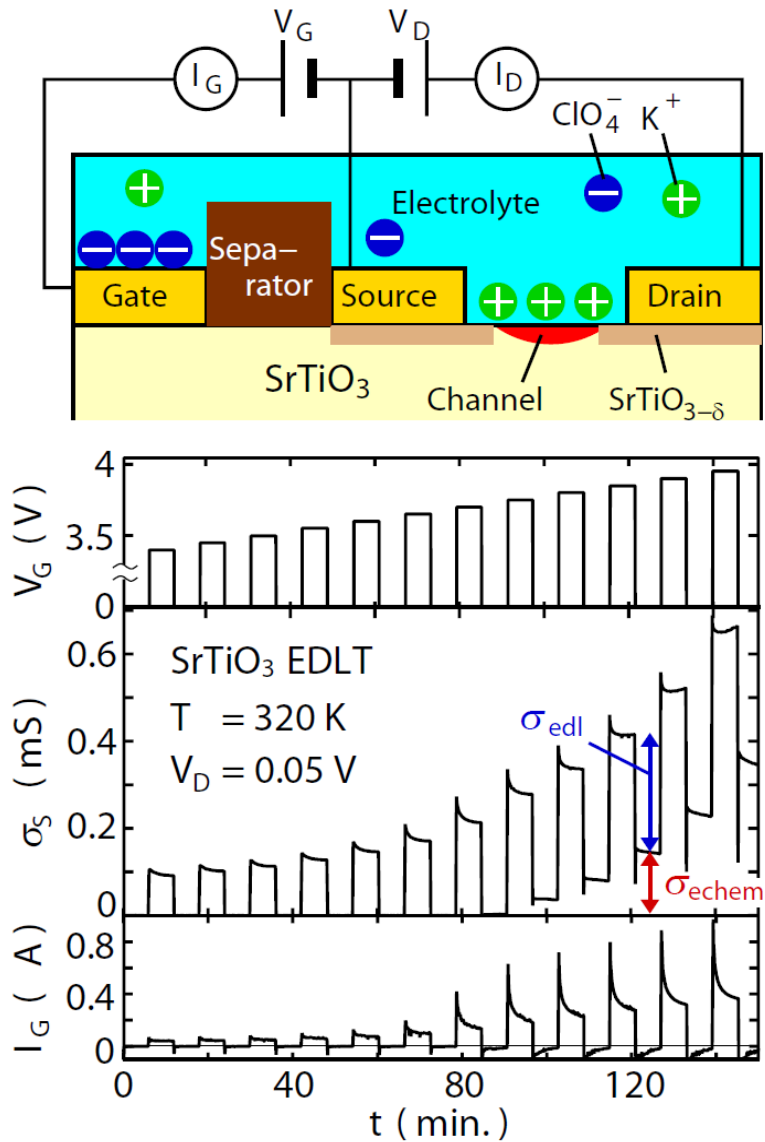

Figure 1: (a) Schematic diagram of an electric double-layer transistor with a channel of $SrTiO_3$. $K^+$ cations (green) and $ClO_4^-$ anions (blue) in the polymer electrolyte are collected on the channel and gate electrode, respectively, and charge carriers are accumulated on the channel (red area). On applying a very high gate bias, this region is damaged. (b) Switching characteristics of a $SrTiO_3$ EDLT at 320 K. Temporal evolutions of gate bias, $V_G$ (top), four-terminal sheet conductance, $\sigma_S$ (middle), and gate leak current, $I_G$ (bottom), are plotted. $\sigma_{edl}$ and $\sigma_{echem}$ indicated by arrows are conductance gains due to the reversible electrostatic charge accumulation and irreversible electrochemical doping, respectively.

**Figure 2**

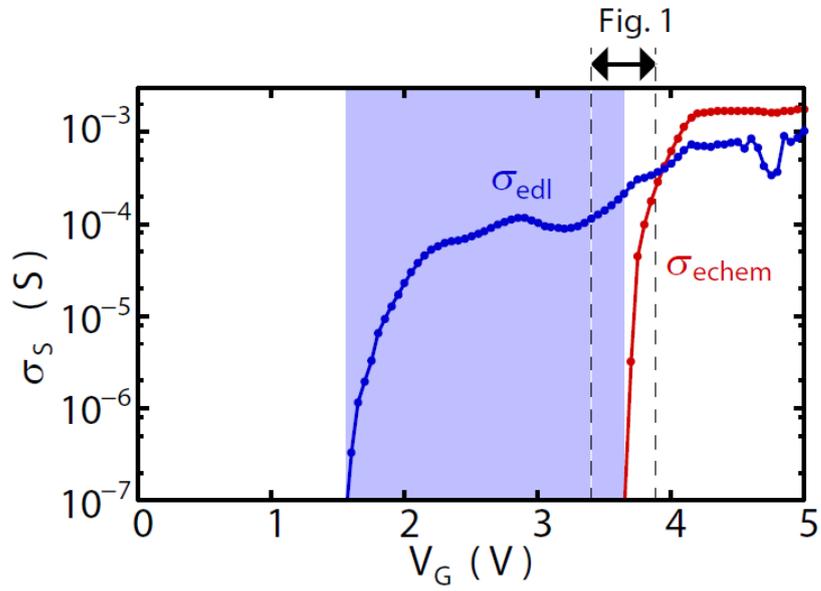

Figure 2: Transfer characteristics of the EDLT at 320 K deduced from the switching characteristics shown in Fig. 1. The blue shadow area corresponds to the $V_G$ range, which gives electrostatic charge accumulation without electrochemical doping. An arrow indicates a $V_G$ range corresponding to Fig. 1.

**Figure 3**

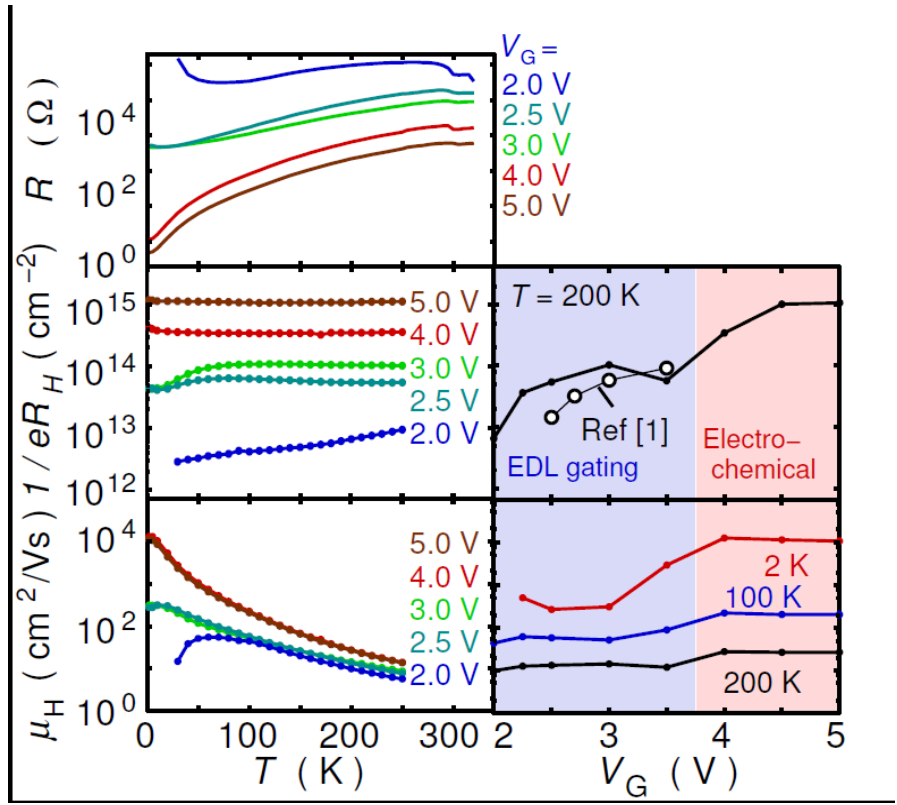

Figure 3: Transport properties of the SrTiO$_3$ channel as a function of temperature for various $V_G$'s (left panel) and as a function of $V_G$ for various temperatures. The top panel shows channel sheet resistance, $R$. The middle and bottom panels show the sheet carrier density, $n_S = 1/eR_H$, and mobility, $\mu_H$, evaluated by Hall-effect measurements. Open symbols in the middle right panel correspond to the data shown in Ref. 1. Blue and red regions in the left panels correspond to the gate bias regions with and without a persistent conduction, respectively.

**Figure 4**

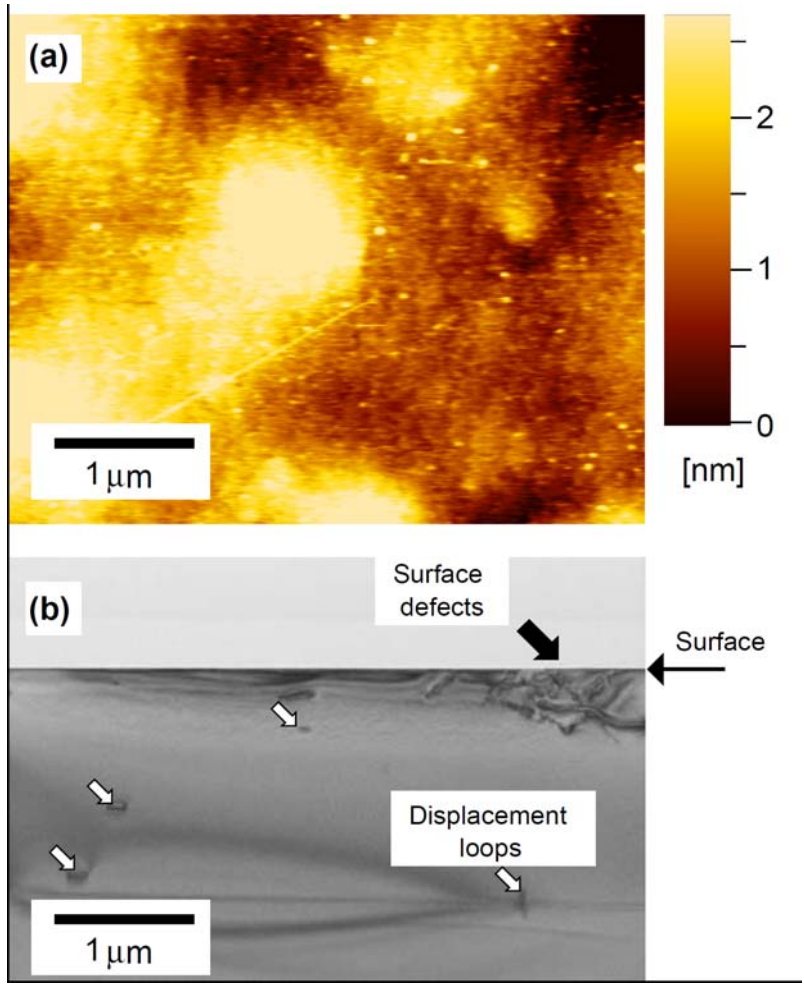

Figure 4: (a) AFM and (b) cross-sectional TEM images after applying $V_G$ of 5 V to induce irreversible conduction path. A black arrow indicates defect structures near the surface, corresponding to an undulated structure in the AFM image. White arrows indicate dislocation loops possibly induced by an oxygen removal. Dark horizontal lines, about 2 μm deep and near the surface of the TEM image are artifacts caused by the sample preparation.

Supplementary information

# Electrostatic charge accumulation vs. electrochemical doping in SrTiO$_3$ electric double layer transistors

**Comparison between a pulse measurement and a sweep measurement**

The repeated application of the gate voltage pulses shown in Fig. 1 could be a possible origin of the electrochemical reactions. In order to rule out this possibility, we also examined the transfer characteristics by a continuous sweeping of the gate bias $V_G$. We measured the transfer characteristics for another device fabricated on the same substrate while $V_G$ was swept at a rate of 5 mV/sec from 0 V to 5 V (black curve) and then from 5 V to 0 V (green curve). These data are superimposed on the Fig. 2 in main part as shown in Fig. S1. The latter curve indicates persistent conduction even after removal of $V_G$. Therefore, the application of a high gate bias commonly induces the electrochemical reaction and the persistent conduction to the SrTiO$_3$ channel.

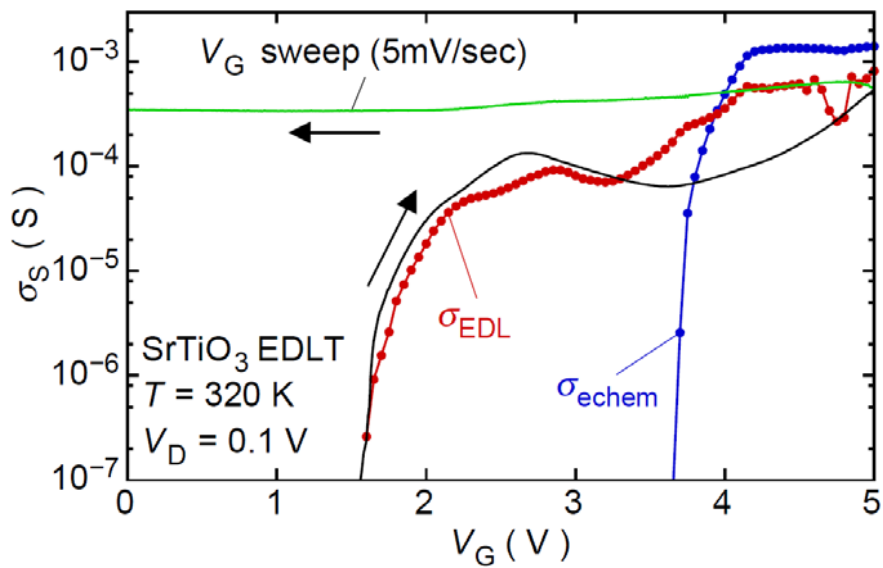

Fig. S1 Transfer characteristics examined by a continuous gate bias sweep. The transfer characteristics in Fig. 2 are also shown for comparison.